\documentclass[pra,amsmath,amssymb,amsfonts,nofootinbib]{revtex4}

\usepackage{bm,graphicx,mathrsfs}

\usepackage{amsmath,bbm}
\usepackage{amsfonts,amssymb}
\usepackage{graphicx}
\usepackage{epsfig}
\usepackage{times}
\usepackage{verbatim}

\newcommand{\be}{\begin{equation}}
\newcommand{\ee}{\end{equation}}
\newcommand{\bq}{\begin{eqnarray}}
\newcommand{\eq}{\end{eqnarray}}
\newcommand{\ba}{\begin{align}}
\newcommand{\ea}{\end{align}}

\newcommand{\1}{\mathbbm{1}}

\newcommand{\ket}[1]{\left | \, #1 \right\rangle}
\newcommand{\bra}[1]{\left \langle #1 \, \right |}

\newcommand{\braket}[2]{\left\langle\, #1\,|\,#2\,\right\rangle}

\newcommand{\proj}[1]{\ket{#1}\bra{#1}}

\newcommand{\avr}[1]{\left \langle#1 \right \rangle}

\newcommand{\tr}[1]{{\rm tr}\left[{#1}\right]}

\newcommand{\un}[1]{{\underline{#1}}}

\newcommand{\raw}{\rightarrow}

\newcommand{\bR}{\mathbbm{R}}

\newcommand{\bC}{\mathbbm{C}}

\newcommand{\cH}{\mathcal{H}}
\newcommand{\cE}{\mathcal{E}}

\newcommand{\cL}{\mathcal{L}}
\newcommand{\cA}{\mathcal{A}}
\newcommand{\cS}{\mathcal{S}}

\newcommand{\cM}{\mathcal M}

\newcommand{\cO}{\mathcal O}

\newcommand{\Var}{{\rm Var}}

\newtheorem{theorem}{Theorem}
\newtheorem{lemma}[theorem]{Lemma}

\newtheorem{proposition}[theorem]{Proposition}
\newtheorem{definition}[theorem]{Definition}

\def\Proof{\noindent\textsc{Proof:}}
\def\proof{\Proof}
\def\qed{\leavevmode\unskip\penalty9999 \hbox{}\nobreak\hfill
     \quad\hbox{\leavevmode  \hbox to.77778em{%
               \hfil\vrule   \vbox to.675em%
               {\hrule width.6em\vfil\hrule}\vrule\hfil}}
     \par\vskip3pt}
    {\hspace*{\fill}$\Box$\vspace{1.5ex}\par}

\newcommand{\Sp}{\,\,\,\,\,\,}

\usepackage{color}

\begin{document}

\title{\sc \large Lower bounds to the spectral gap of Davies generators}

\author{Kristan Temme\\}
\address{Center for Theoretical Physics, Massachusetts Institute of Technology, Cambridge, MA 02139, USA}
\date{\today}

\begin{abstract}
We construct lower bounds to the spectral gap of a family of Lindblad generators known as Davies maps. These maps describe the thermalization of quantum systems weakly coupled to a heat bath. The steady state of these systems is given by the Gibbs distribution with respect to the system Hamiltonian. The bounds can be evaluated explicitly, when the eigenbasis and the spectrum of the Hamiltonian is known. A crucial assumption is that the spectrum of the Hamiltonian is non-degenerate. Furthermore, we provide a counterexample to the conjecture, that the convergence rate is always determined by the gap of the associated Pauli master equation. We conclude, that the full dynamics of the Lindblad generator has to be considered. Finally, we present several physical example systems for which the bound to the spectral gap is evaluated.
\end{abstract}

\maketitle

\section{Introduction}\label{sec:Intro}

A particular class of Liovillians \cite{Lindblad,Breuer}, which describes the thermalization of a quantum mechanical subject to thermal noise, is known as Davies generators \cite{Davies,Davies2}. This class of Liouvillians  describes the dissipative dynamics resulting as the weak (or singular) coupling limit from a joined Hamiltonian evolution of a system coupled to a large heat bath. The weak coupling limit permits to consider only the reduced dissipative dynamics, which gives rise to a Markovian semi-group generated by the aforementioned Davies Liovillian. This generator retains the information of the bath temperature $\beta$ and converges to the Gibbs distribution of the system, constructed from the system Hamiltonian $H$.  The full Hamiltonian of the joined system is given by the sum of the system Hamiltonain, the bath Hamiltonian $H_B$ and a weak interaction $H_{I}$, 
\be
	H_{tot} = H + H_{B} + H_{I} \Sp \mbox{where,} \Sp H_{I} = \sum_{\alpha} S^{\alpha} \otimes B^{\alpha}.
\ee
Note, that both the system's coupling operators $S^\alpha$ as well as the bath operators $ B^{\alpha}$ are Hermitian. Within an appropriate limit the system's evolution can be described  by a Davies generator $ \cL_\beta(f)$. See \cite{Davies,Davies2,Spohn} for a clear derivation. Here, we assume that  such a generator is  already given in is in the canonical form.  We discuss the properties of this generator directly, and are not concerned with the actual derivation. We assume, that we are able to diagonalize the system Hamiltonian $H$ and  can write 
\be
	H = \sum_k \epsilon_k \proj{k},
\ee
where $\epsilon_k$ are the eigenvalues to the eigenvectors labeled by $\ket{k}$.  We choose the Heisenberg picture as a convention so that the generator describes the evolution of observables.  The canonical form of the Davies generator is given by   
\be\label{thermalLio}
\cL_\beta(f)=  i[H,f] + \sum_{\omega,\alpha}\cL_{\omega,\alpha}(f).
\ee
 The individual summands are 
\bq 
\cL_{\omega,\alpha} (f) = G^\alpha(\omega)\left({S^\alpha}^\dag(\omega)fS^\alpha(\omega) - \frac{1}{2} \{{S^\alpha}^\dag(\omega)S^\alpha(\omega),f\} \right),
\eq 
where variable $\omega$ refers to the so-called Bohr frequencies of the system Hamiltonian, i.e. energy differences $\omega = \epsilon_i - \epsilon_j$, and the index $\alpha$ enumerates the coupling operators to the environment. The functions $G^\alpha(\omega)$ correspond to the Fourier transform of the two point correlation functions of the environment, and are bounded \cite{Breuer,Spohn}. These functions depend on the bath operators as well as the thermal state of the bath and encode the equilibrium temperature. The  Lindblad operators are given by the Fourier components of the coupling operators $S^\alpha$ which evolve according to the system Hamiltonian 
\be
	e^{iH t} S^\alpha e^{-i H t } = \sum_\omega S^\alpha(\omega) e^{i \omega t}.
\ee
The operators $S^\alpha(\omega)$  induce transitions between the eigenvectors of $H$ with energy $E$ to eigenvectors of $H$ with energy $E+\omega$, and hence act as quantum jump operators, which transfer energy $\omega$ from the system to the bath. A direct evaluation shows, that the operators $S^\alpha(\omega)$ are of  the form
\be\label{Davi-lind}
	S^\alpha(\omega) = \sum_{km} S^\alpha_{km}(\omega) \ket{k}\bra{m},
\ee
where we have  defined
\bq
S^{\alpha}_{km}(\omega) = \delta\left[ \epsilon_k - \epsilon_m - \omega \right]  S_{km}^\alpha, \Sp \mbox{with} \Sp \delta[x] = \left\{\begin{array}{l} 1 \Sp \mbox{if} \Sp x = 0 \\ 0 \Sp \mbox{else}
\end{array}\right.,
\eq
with $S_{km}^\alpha = \bra{k}S^\alpha\ket{m}$. Under certain conditions \cite{Spohn} on the operators $S^\alpha(\omega)$  the thermal map can be seen to have a unique full-ranked stationary state which is given by $\sigma \propto e^{-\beta H}$, where $\beta$ is the inverse temperature of the heat bath. The following useful relations hold for any $ \alpha$ and $\omega$: 
\bq 
G^\alpha(-\omega)&=&e^{-\beta \omega}G^\alpha(\omega)\label{DBDavies1}\\ 
\sigma S^\alpha(\omega)&=&e^{\beta \omega}S^\alpha(\omega)\sigma \label{DBDavies2}.
\eq 
The condition (\ref{DBDavies1}) for the functions $G^\alpha(\omega)$ is often referred to as KMS condition \cite{kubo1957statistical,martin1959theory} and ensures together with (\ref{DBDavies2}) the reversibly (c.f. Defintion \ref{DetailedBalance}) of the generator $\cL_\beta$ \cite{db1,Majewski,Streater,alicki1976detailed}, as can be verified easily.  Reversibility ensures furthermore, that the generator has a real spectrum, which is contained in $(-\infty,0]$. We will elaborate on this in a later section.\\

In this article we are concerned with the derivation of general lower bounds to the spectral gap $\lambda$ of Davies generators.  Under the assumption that the spectrum as well as the matrix elements in the energy eigenbasis of the coupling operators $S^\alpha$ are known, we provide a formula for a lower bound to the spectral gap, which can be evaluated on a case by case basis. We make the assumption that the system Hamiltonian $H$ has only non-degenerate eigenvalues. 

Several techniques exist for bounding the spectral gap of classical Laplacians or Markov processes \cite{levin2009markov}, such as the canonical path lemma or other geometrical bounds \cite{diaconis1991geometric,fill1991eigenvalue}, which rely on the graph representation of the Markov processes. These estimates can give rise to exponentially decaying bounds on the convergence of the Markov process to its fixed point measured in total variational distance. We are interested in deriving similar bounds for the convergence of quantum Markovian semi-groups measured in trace- or Schatten 1 norm. The bounds for the family of Davies generators presented here can be seen as a generalization of the classical techniques to quantum mechanical semi-groups. A lower bound to the spectral gap of the Davies generator for a particular system was already derived in \cite{Alicki}, but such bounds on the spectrum are in general rare and difficult to obtain. We expect that the bounds derived here will find applications in the estimation of thermalization or decoherence times \cite{Breuer,Schlosshauer} or lifetime  of quantum memories \cite{terhal,Alicki}. Furthermore, convergence bounds are needed in the stability analysis of quantum mechanical semi-groups \cite{cubitt2013stability} and can be used to provide estimates of the correlation length in the steady state of the system \cite{kastoryano2013rapid}.\\

In the following section \ref{sec:Convergence} we establish the formal background and discuss means to bound the convergence time, also often referred to as mixing time, to the steady state. We state some elementary lemmata characterizing the spectral gap of the Liovillian.  We then proceed in section \ref{sec:Dirichlet} to discuss the particular form of the Dirichlet form of the Davies generator and show that it can be written as the quadratic form of a block diagonal matrix. This direct sum decomposition facilitates the derivation of the lower bound greatly. As was already observed by several authors \cite{Davies,Davies2, alicki1977markov, roga2010davies}, one block of the generator corresponds to the so called Pauli - master equation, which is a rate equation for the populations in the eigenbasis of the Hamiltonain. It would be tempting to conjecture that it is in fact only the spectrum of this block that determines the convergence. A simple counter example will show that this is not true, and that one in fact has to consider the spectrum of all the blocks in this decomposition. In the subsequent section \ref{sec:lowerBounds} we then proceed to derive lower bounds to the spectral gap by bounding the lowest non-trivial eigenvalue in each block. We then present several examples to which the spectral bounds are applied in section \ref{sec:Applicaitons}. \\

Throughout this article we only consider operators acting on finite dimensional Hilbert spaces with $\dim{\cH} = d$, which are isomorphic to the algebra of $d$-dimensional complex matrices $\cM_d \cong \bC^{d\times d}$. We will write for the vector representation of a matrix  $\ket{f} = \sum_{ab} f_{ab} \ket{a\overline{b}} = f \otimes \1 \ket{\Omega}$, where $\ket{\Omega} = \sum_k \ket{kk}$ denotes the unnormalized maximally entangled state. The matrix space $\cM_d$ is equipped with the canonical Hilbert Schmidt scalar product $\braket{g}{f} = \tr{g^\dag f}$ and it can be verified easily that the following identity $\ket{AXB} = A\otimes B^T \ket{X}$ holds for the vectorization. Here $B^T$ denotes the transpose with respect to the chosen basis. We will furthermore consider operators $\cO : \cM_d \raw \cM_d$ mapping matrices to matrices. The matrix representation of the operator $\cO$ on $\bC^{d\times d}$ will be written as $\hat{\cO}$, where we assume that the entries $[\hat{\cO}]_{ab} = \tr{F^\dag_a\cO(F_b)}$ can be computed from a suitable matrix basis $\{F_a\}_{a=1,\ldots,d^2}$.  We denote the set of $d$-dimensional Hermitian operators  by $\cA_d=\{X\in\cM_d,X=X^\dag\}$, and write for the subset of positive definite operators $\cA^+_d=\{X \in \cA_d, X > 0\}$. The set of states is denoted by $\cS_d=\{X \in \cA_d,X\geq0,\tr{X}=1\}$, and we will write for states of full rank $\cS_d^+$.  The Hermitian conjugate with respect to the canonical scalar product on $\bC^d$ will be written as $f^\dag$, whereas the conjugate of the operator $\cO$ with respect to the Hilbert Schmidt scalar product is denoted by  $\cO^*$. Note, that these two notations coincide, when we consider the matrix representation $\hat{\cO}$. The complex conjugate  $x \in \bC$ is denoted by $\overline{x}$.

%
% Formal background
%
\section{Formal background and mixing time bounds}\label{sec:Convergence}

The spectral properties of the generator (\ref{DBDavies1}) can best be understood, when working with an inner product that  is weighted with respect to some full rank reference state $\sigma \in \cS_d^+$.  This reference state is typically chosen as the fixed point of the Lioviallian, i.e. the Gibbs state. The weighting can be expressed in terms of a map acting on elements  $f \in \cA_d$ by writing $\Gamma_\sigma(f)  =  \sigma^{1/2} f \sigma^{1/2}$. Note, that this choice of $\Gamma_\sigma$ is not unique due to the non-commutativity of $f$ and $\sigma$. In fact, there exists an entire family of possible maps, which all stem from monotone Riemanian metrics \cite{chi2,Petz1,Petz2,RuskaiRiem}. For reasons of simplicity we will work with the particular choice stated above. We will furthermore denote the eigenvalues of the linear map $\Gamma_\sigma$ by $\sigma(a,b) = \sqrt{\sigma_a \sigma_b}$. The corresponding eigenoperators are given by $\ket{a}\bra{b}$, which are just the matrix units in the eigenbasis of $\sigma = \sum_a \sigma_a \proj{a}$. With this operator at hand, we can proceed to define the weighted inner product, the corresponding variance and most importantly the Dirichlet form associated with the Liovillian generator $\cL$. 

\begin{definition}
Given a Liouvillian $\cL:\cM_d\rightarrow\cM_d$ with unique full rank stationary state $\sigma$ and associated map $\Gamma_\sigma(f)  =  \sigma^{1/2} f \sigma^{1/2}$,  we define 
\begin{enumerate}
\item
The $\sigma$-weighted non-commutative \textbf{inner product}  by 
\be\label{inner-prod}
\avr{f,g}_\sigma = \tr{\Gamma_\sigma(f)g} = \tr{\sigma^{1/2} f \sigma^{1/2} g}.
\ee

\item
The weighted \textbf{variance} defined as 
\be\label{weighted-variance}
\Var_\sigma(g,g) = \tr{\Gamma_\sigma(g) g} - \tr{\Gamma_\sigma(g)}^2. 
\ee 

\item
The \textbf{Dirichlet form} of the generator $\cL$: 
\be \cE(f,f) = -\avr{f, \cL(f)}_\sigma = - \tr{\Gamma_\sigma(f) \cL(f)}. \label{def:Dirichlet} 
\ee 
\end{enumerate}
\end{definition} 

These quantities give convenient access to the spectral properties of the Liovillian $\cL$ as is outlined in the reference \cite{chi2} in greater detail. We will only discuss the most relevant aspects here.  In general an exponential convergence bound on the trace distance can be given in terms of a constant that is related to the lowest non-vanishing eigenvalue of a  weighted, additive, symmetrization of the generator  $\cL$. In the particular case of the Davies generator, this constant coincides with the spectral gap  of the original Liovillian.  This fact is a consequence of the reversibility of the Davies generator. 

\begin{definition}[Detailed balance]\label{DetailedBalance} We say a Liouvillian $\cL:\cM_d\rightarrow\cM_d$ satisfies \textbf{detailed balance}
 (or is \textbf{reversible}) with respect to the state $\sigma\in\cS_d^+$, if $\Gamma_\sigma \circ \cL = \cL^* \circ \Gamma_\sigma$.
\end{definition}

It can be verified easily that the Davies generator, which fulfills the KMS condition, is reversible \cite{db1,Majewski,Streater,alicki1976detailed} with respect to the Gibbs distribution as was already shown in the seminal work by Davies \cite{Davies,Davies2}. It is furthermore possible to see, that the generator is also reversible with respect to the definition as given above with respect to $\Gamma_\sigma$. This immediately implies two things. First, we observe that a reversible Lioviallian becomes self - adjoined with respect to the weighted inner product as defined in (\ref{inner-prod}). This in turn implies that the spectrum of $\cL$ is real. Second, as can be verified easily, reversibility ensures that the state $\sigma$ is a fixed point of the Liovillian \cite{chi2}. We are now ready to find a convenient variational expression for the spectral gap of the Davies generator. The following lemma was already proved in \cite{chi2}.

\begin{lemma} The \textbf{spectral gap} of a primitive Liouvilian $\cL:\cM_d\rightarrow\cM_d$ with stationary state $\sigma$ is the largest constant  $\lambda$, so that 
\bq \label{spectGapVar}
	\lambda \Var_\sigma(g,g)  \leq \cE(g,g).
\eq  
for all $g \in \cA_d$.
\end{lemma}

\proof{ We start by defining a map $Q$ which is self adjoined with respect to the canonical Hilbert Schmidt scalar product $\braket{g}{f} = \tr{g^\dag f}$.  Let $Q = 1/2(\Gamma_\sigma^{1/2} \circ \cL^* \circ \Gamma_\sigma^{-1/2} + \Gamma_\sigma^{-1/2}\circ \cL \circ \Gamma_\sigma^{1/2})$. Note, furthermore, that if $\cL$ is reversible we have that $Q$ is related to $\cL$ by a similarity transformation.  The map $Q$ has an eigenvector $\sigma^{1/2}$ that corresponds to the eigenvalue $\lambda_0 = 0$. The spectral gap can therefore be expressed as \cite{Bhatia}
\be
	\lambda = \min_{f \in \cA_d; \tr{\sqrt{\sigma}f}  = 0} \frac{-\bra{f}Q\ket{f}}{\braket{f}{f}}.
\ee
This can be rewritten as 
\be
	\lambda = \min_{f \in \cA_d; \tr{\sqrt{\sigma}f} = 0} \frac{-\bra{f}Q\ket{f}}{\braket{f}{f} - \braket{\sqrt{\sigma}}{f}^2},
\ee
since the constraint ensures that $\braket{\sqrt{\sigma}}{f} = 0$. Note, however, that now both the nominator, as well as the denominator are invariant under transformations of the form $f \raw f + c\sqrt{\sigma}$, where $c$ is some constant. This in turn implies that we may drop the aforementioned constraint for $f$ since for every $f$ which is not orthogonal to $\sigma^{1/2}$ we may find a $c$ and an associated transformation which makes this $f$ orthogonal. Hence, when we substitute $f = \Gamma^{1/2}_\sigma(g)$, we are left with $\lambda = \min_{g \in \cA_d} \cE(g,g)/\Var_\sigma(g,g)$. \qed}

\vspace{0.3cm}
We are now ready to discuss the convergence behavior of the Davies generator. In order to quantify the convergence to the steady state, we need to choose an appropriate norm. The convergence is most often estimated in trace norm, $\| A \|_{tr} = \tr{|A|}$ since it possesses a clear operational interpretation \cite{Fuchs}. However, it is much more convenient to work with other distance measures when deriving upper bounds to the convergence time. We will work with  the $\chi^2$- divergence defined in \cite{chi2}. This divergence is related to the trace norm through the bound

\bq\label{TraceBounds}
\left\| \rho - \sigma \right \|_{tr}^2  \leq  \chi^2(\rho,\sigma) = \tr{(\rho-\sigma)\Gamma_\sigma^{-1}(\rho-\sigma)}.
\eq

The variance  $\Var_\sigma(g,g)$ (\ref{weighted-variance}) coincides with the $\chi^2$-divergence for a particular choice of $g$. We have for $\tilde{g} = \Gamma^{-1}_\sigma(\rho - \sigma)$ the equality $\Var_\sigma(\tilde{g},\tilde{g}) = \chi^2(\rho,\sigma)$. If one therefore bounds the evolution of the variance for all values of $g$, a convergence bound for the $\chi^2$-divergence can be given. This bound in turn implies a convergence bound for the trace norm due to the inequality (\ref{TraceBounds}). 

\begin{theorem}\label{Mixingtimebound}
Let $\cL:\cM_d\rightarrow\cM_d$ be a Liouvillian with stationary state $\sigma$ and spectral gap $\lambda$ Then the following trace norm convergence bound holds: 
\be
	\left\| \rho_t - \sigma \right \|_{tr} \leq \sqrt{\sigma_{min}^{-1}}e^{-\lambda t}.
\ee  
Here $\sigma_{\min}$ denotes the smallest eigenvalue of the stationary state $\sigma$.
\end{theorem}

\proof{ Let us for simplicity only consider the case when $\cL$ is detailed balanced. The result also holds when this is not the case, but the proof requires some notational overhead \cite{chi2}. Define $g_t = \Gamma^{-1}(\rho_t - \sigma)$, where $\rho_t$ evolves according to the equation $\dot{\rho}_t = \cL^*(\rho_t)$. When $\cL$ is reversible, it can be verified that this implies an evolution for $g_t$ which is $\dot{g}_t = \cL(g_t)$. The time derivative of the variance yields
\be
\partial_t \Var_\sigma(g_t,g_t) = -2\cE(g_t,g_t) \leq -2\lambda \Var_\sigma(g_t,g_t), 
\ee
where the last inequality is due to the variational characterization of the gap $\lambda$. Integration of this differential inequality gives us the bound $ \Var_\sigma(g_t,g_t) \leq e^{-2\lambda t}\Var_\sigma(g_0,g_0)$. This in turn, by back substitution of $g_t$ and by virtue of (\ref{TraceBounds}), yields the bound on the trace norm
\be
\left\| \rho - \sigma \right \|_{tr}^2  \leq  e^{-2\lambda t} \chi^2(\rho_0,\sigma).
\ee
An optimization over all input states gives rise to the upper bound $\chi^2(\rho_0,\sigma) \leq \sigma_{min}^{-1}$ which can be attained when $\rho_0$ is the projector onto the smallest eigenvalue of $\sigma$. \qed}

\vspace{0.3cm}
We can define the mixing time $t_{mix}$ as the time the semi-goup $\cL$ needs to be $\epsilon$-close to its stationary distribution. Hence, we define 
\be
	t_{mix}(\epsilon) = \left\{ t \;\;\big {|}\;\;  t' > t  \Sp \mbox{we have} \Sp \|e^{\cL t}(\rho_0) - \sigma\|_{tr} \leq \epsilon \;\;\forall \rho_0 \right\}.
\ee
The convergence result of theorem \ref{Mixingtimebound}, provides a simple upper bound on the mixing time. One can easily rearrange the upper bound to find that we can choose 
\be\label{bound-on-t}
	t_{mix}(\epsilon) \leq \frac{\log(\sigma_{min}^{-1}/\epsilon^2)}{2\lambda}.
\ee

\subsection{The spectral gap and  the support number of a matrix pencil}

Due to lemma \ref{spectGapVar}, it is clear that  the problem of finding good lower bounds to the spectral gap can be rephrased as the problem of finding a constant $\lambda$, so that the inequality $\lambda \Var_{\sigma}(g,g) \leq \cE(g,g)$, is satisfied for all  $g \in \cA_d$.  A controlled approach to finding lower bounds to $\lambda$ in this inequality is  by support theory \cite{boman2003support,bern2001support}. First developed to construct good preconditioners for linear systems \cite{spielman2003solving}, it was also used to improve on the spectral gap bounds of graph Laplacians. Here, we will only briefly state some of the results which are immediately relevant to us. 

As we will see later, the inequality (\ref{spectGapVar}) can be expressed in terms of a matrix inequality. We make use of the vectorization of $f \in \cA_d$ through $\ket{f} = f \otimes \1 \ket{\Omega}$ so both the quadratic forms can be written as $\Var_{\sigma}(f,f) = \bra{f} \hat{V}_\sigma \ket{f}$ and $\cE(f,f) = \bra{f} \hat{\cE} \ket{f}$.  The matrices $ \hat{V}_\sigma$ and $\hat{\cE}$ will be introduced in section \ref{sec:Dirichlet}. The inequality (\ref{spectGapVar}) can therefore be expressed as 
\be
	\tau \hat{\cE} - \hat{V}_\sigma \geq 0,
\ee
when we define $\tau = \lambda^{-1}$.  The problem  reduces to finding the smallest $\tau$  for which this matrix inequality is satisfied. Hereby we mean, that the resulting matrix is positive semi definite. Any upper bound on $\tau$ will immediately constitute a lower bound on the spectral gap $\lambda$. This constant $\tau$ is referred to as the \emph{support number}. We give the following definition 

\begin{definition}
The support number $\tau$ of the matrix pair $(A,B)$  with $A,B \in \cM_d(\bC)$ is defined as 
\be
	\tau(A,B) = \min\left\{ t \in \bR \;\; | \;\; rB - A \geq 0 \Sp \forall r \geq t \right\}.
\ee
\end{definition}

Note, that the support of a matrix pair is well defined even for singular matrices, as long as $\ker(B) \subset \ker(A)$. This will be the case for the matrix pencil $(\hat{V}_\sigma,\hat{\cE})$, since $\cL$ is assumed to be primitive and $\sigma$ is its only fixed point.  Note, furthermore, that we only consider matrices which are Hermitian and positive semi-definite. In general, however, support theory is not restricted to this setting.

A very simple but useful lemma of support theory is the splitting lemma. It provides a method for braking up the big problem of finding a bound on $\tau$ into smaller subproblems.
 
\begin{lemma}
If for  positive semi definite matrices $A,B$ there is a splitting $A = \sum_{i=1}^q A_i$ and $B = \sum_{i=1}^q B_i$, where $A_i,B_i \geq 0$, then 
\be
	\tau(A,B) \leq \max_i \tau(A_i,B_i).
\ee
If, furthermore we have that $A = \oplus_i A_i$ and $B = \oplus_i B_i$, equality holds. 
\end{lemma}
\proof{ This bound follows immediately from the variational characterization of eigenvalues \cite{Bhatia}. \qed}

A further tool to bound the support number was developed in \cite{boman2003support,chen2005obtaining}.  It is possible to express the support number as the constrained minimization over a certain matrix factorization. Hence, any factorization that satisfies the constraints gives rise to a valid upper bound on the support number. This is expressed in the following lemma. For completeness we will repeat the proof here.

\begin{lemma}\label{lem:W-bound}
Let $A,B$ be positive semi-definite with a decomposition $A = U U^\dag$  and $B = V V^\dag$. then the support of $(A,B)$  is given by 
\be
	\tau(A,B) = \min_{W} \| W \|_{2\raw2}^2 \Sp \mbox{subject to} \;\; V W =U.
\ee
\end{lemma}

\proof{ Note, that due to Silvester's law of inertia \cite{Bhatia}, we have that for any $\tau B - A \geq 0$ the conjugation $S (\tau B -  A)S^\dag \geq 0$ is also positive semi-definite. Hence, we have that  $\tau(A,B) \geq \tau(SAS^\dag,SBS^\dag)$ for any matrix $S$. In particular, if $\mbox{ker}(S) \subset \mbox{ker}(A)$ and $\mbox{ker}(S) \subset \mbox{ker}(B)$, we have that $\tau(A,B) = \tau(SAS^\dag,SBS^\dag)$ due to the variational characterization of eigenvalues. Suppose we have a decomposition as stated in the lemma, then
\be
	\tau B - A = \tau V V^\dag - U U^\dag = V\left(\tau \1 - W W^\dag \right)V^\dag,
\ee
and thus $\tau(A,B) \leq  \min_{W} \| W \|_{2\raw2}^2$.  We see that the minimum can be attained for $W = V^\#U$, where $V^\#$ denotes the Penrose inverse of $V$.\\ \qed}

Hence, any matrix $W$ which satisfies the constraints yields an upper bound to the support number.  The direct evaluation of the  $2\raw2$ norm does at first appear to be just as challenging as the original problem. However, since we are only trying to find upper bounds on $\tau(A,B)$ suitable norm inequalities will suffice. In particular, we have the well known inequalities \cite{Bhatia}
\bq
&& \| W \|_{2\raw2}^2 \leq \| W \|_F^2, \label{W:frob}\\
&& \| W \|_{2\raw2}^2 \leq \| W \|_1\| W \|_\infty.  \label{W:conj_dil}
\eq
These norm bounds give rise to some of the well known spectral gap bounds for graph Laplacians for a suitably chosen decomposition $(V,U,W)$. For instance, the congestion - dilation lemma \cite{spielman2003solving} can be obtained, when $W$ corresponds to the embedding of the associated graph into the complete graph and suitable norm bounds \cite{chen2005obtaining} are applied. 

\section{The Dirichlet form of the Davies Generator}\label{sec:Dirichlet}

Before we proceed let us make the following assumption. We assume, that the system Hamiltonian $H = \sum_k \epsilon_k \proj{k}$ has a spectrum that is non-degenerate. This is a condition which typically holds for a Hamiltonian with no symmetries. The Dirichlet form of the Davies generator is given by the sum of individual forms for different values of $\alpha,\omega$. We can write for every  $f \in \cA_d$
\be
	\cE(f,f) = - i \tr{\Gamma_\sigma(f)\left[H,f\right]} - \sum_{\alpha, \omega} \tr{\Gamma_\sigma(f)\cL_{\alpha,\omega}(f)}.
\ee
We observe that due to the special form of the fixed point $\sigma \propto \exp(-\beta H)$, the Hamiltonian $H$ commutes with any power of fixed point and we can easily see that the first summand $\tr{\Gamma_\sigma(f)\left[H,f\right]} = 0$ vanishes for all $f$. The Hamiltonian therefore does not contribute directly to the spectral gap of the Davies generator and we will ignore this summand in the Dirichlet form from now on.\\

Before we proceed, we introduce some new notation. If we denote by $\nu = \epsilon_n - \epsilon_m$ a particular Bohr frequency, which is determined by some pair  $\epsilon_n,\epsilon_m$, then we denote the set of all energy pairs which give rise to the same Bohr frequency by
\be
	\hat{\nu} = \Big{\{ }\;\; (n_1,n_2) \;\;\Big{|}  \;\; n_i = 0, \ldots, d\;\; \mbox{with} \;\; \nu = \epsilon_{n_2} - \epsilon_{n_1} \;\; \Big{\}}.
\ee 
Furthermore we denote by $\underline{n} \in \hat{\nu}$ the tuple $(n_1,n_2)$ of energy labels in $\hat{\nu}$.

\begin{lemma}\label{non-degForm}
Let $\cL_\beta$ denote the thermal Liovillian, as defined in (\ref{thermalLio}), with the non-degenerate Hamiltonian $H = \sum_k \epsilon_k \proj{k}$ where furthermore $\{S^\alpha\}$ are the coupling operators, then the Dirichlet form (\ref{def:Dirichlet}) assumes the form
\be
	\cE(f,f) = \bra{f} \hat{\cE} \ket{f},
\ee
where $\ket{f} \equiv f\otimes \1 \ket{\Omega}$ is the vectorization of the matrix $f \in \cM_d$. The Dirichlet matrix $\hat{\cE}$ is block diagonal 
\be\label{blockE}
	\hat{\cE} =  \bigoplus_{\nu}  \hat{\cE}^\nu,
\ee
where the direct sum is taken over all Bohr frequencies $\nu$ and each $\hat{\cE}^\nu$ is given by
\bq
 \hat{\cE}^\nu &= &  \sum_{\un{m} \in \hat{\nu};\alpha} \frac{1}{2}\sum_{i=1}^2 \bra{m_i}S^\alpha G^\alpha_{m_i} S^\alpha \ket{m_i}  \sigma(m_1,m_2) \; \;  \proj{\un{m}} \nonumber \\ & -& \sum_{\un{m}\un{l} \in \hat{\nu};\alpha} G^\alpha(\epsilon_{l_1} - \epsilon_{m_1})\sigma(m_1,m_2) \; \overline{S}^\alpha_{l_2, m_2} S^\alpha_{l_1,m_1} \; \ket{\un{l}}\bra{\un{m}} \label{junk}.
 \eq
 Here, we have defined the matrix $G^\alpha_m \equiv \sum_k G^\alpha(\epsilon_k - \epsilon_m) \proj{k}$, and $\ket{\underline{m}} = \ket{m_1\overline{m}_2}$. We denote by  $ S^\alpha_{l,m}$, the matrix elements of $S^\alpha$ in the eigenbasis of H.
\end{lemma}

\vspace{0.5cm}
\Proof{ For the Davies generator defined in (\ref{thermalLio}), the Dirchichlet form (\ref{def:Dirichlet}) is given by the following expression: For any $f \in \cM_d$, we have
\bq
\cE(f,f) = \sum_{\omega,\alpha}G^\alpha(\omega)\left(\frac{1}{2}\tr{\Gamma_\sigma(f^\dagger)\{{S^\alpha}^\dag(\omega)S^\alpha(\omega),f\}} -\tr{\Gamma_\sigma(f^\dagger){S^\alpha}^\dag(\omega)fS^\alpha(\omega)} \right).
\eq
Note, that we can always write $\tr{AB} = \bra{\Omega} A \otimes B^T \ket{\Omega}$, so that $ \cE(f,f) = \bra{f}\hat{\cE}\ket{f} $. If we define $\Pi_m = \proj{m}$ as the projector onto the eigenstate with eigenvalue $\epsilon_m$, we can write for $\Gamma_\sigma(f) = \sum_{mn}\sigma(m,n)\Pi_m f \Pi_n$. We then have that the Dirichlet matrix is given by
\bq
\hat{\cE} &=& \sum_{\omega,\alpha}\sum_{nm} G^\alpha(\omega) \frac{\sigma(n,m)}{2}\left(\Pi_m{S^\alpha}^\dag(\omega)S^\alpha(\omega)\Pi_m \otimes \Pi_n^T + \Pi_m \otimes (\Pi_m{S^\alpha}^\dag(\omega)S^\alpha(\omega)\Pi_m )^T\right) \nonumber \\
&-&\sum_{\omega,\alpha}\sum_{nm} G^\alpha(\omega) \sigma(n,m) \Pi_m {S^\alpha}^\dag(\omega) \otimes \Pi_n^T {S^\alpha}^T(\omega)
\eq
Now,  with the matrix $G^\alpha_m \equiv \sum_k G^\alpha(\epsilon_k - \epsilon_m) \proj{k}$ and the definition of the Lindblad operators $S^\alpha(\omega)$ as given in (\ref{Davi-lind}) we can compute the sum over the Bohr frequencies $\omega = \epsilon_n - \epsilon_m$. We are then left with
\bq\label{blockDeg}
\hat{\cE} &=& \sum_{nm;\alpha}\sigma(n,m) \frac{1}{2} \left(\Pi_mS^\alpha G^\alpha_mS^\alpha\Pi_m \otimes \Pi_n^T + \Pi_m \otimes (\Pi_n S^\alpha G^\alpha_nS^\alpha\Pi_n)^T\right) 
\nonumber \\ &-&\sum_{nm;\alpha}\sum_{lk} G^\alpha(\epsilon_n - \epsilon_m) \sigma(m,l)\; \delta[\epsilon_n - \epsilon_m -\epsilon_k + \epsilon_l]
\; \Pi_m S^\alpha \Pi_n \otimes (\Pi_k S^\alpha \Pi_l)^T.
\eq
We see that the first summand in this expression is diagonal in the basis states of the form $\ket{m\overline{n}}$. The second term vanishes, whenever $\epsilon_n - \epsilon_m -\epsilon_k + \epsilon_l \neq 0$, so we only have contributions whenever 
\be
	\epsilon_m - \epsilon_l = \epsilon_n - \epsilon_k \equiv \nu.
\ee
This difference $\nu$ is just another Bohr frequency, which can be either positive or negative. However, these frequencies do not pair the old labels, which gave rise to the $\omega$.  If we now introduce the new labels $(m_1,m_2) = (m,l)$ and $(l_1,l_2) = (n,k)$ and express the sums as sums over these tuples, i.e. sums over $\underline{m},\underline{l} \in \hat{\nu}$, we see that for different values of $\nu$, the resulting matrices do not have joint support. That is, they become block diagonal, where each block corresponds to a different $\nu = \epsilon_0 - \epsilon_d,\dots,0,\ldots\epsilon_d - \epsilon_0$, if we assume an ordering $\epsilon_0 < \epsilon_1\ldots<\epsilon_d$. Hence, $\hat{\cE}$ can be written as a direct sum of matrices  labeled  by $\nu$, as in (\ref{blockE}). \qed}

\vspace{0.3cm}
The matrices $\hat{\cE}^\nu$ are all Hermitian and one furthermore observes that the matrix entries of $\hat{\cE}^{-\nu}$  are related to those of ${\hat{\cE}^\nu}$ by complex conjugation. Hence for all $\nu \neq 0$ the matrices come in pairs and it therefore suffices to focus only on the matrices with $\nu \geq 0$, since the spectra of the matrices $\hat{\cE}^{-\nu}$ and ${\hat{\cE}^\nu}$ coincide. A possible interpretation of the matrices for different  $\nu$  can be given as follows:
 
For Hamiltonians with non-degenerate eigenvalues, it was already observed in  \cite{Davies,Davies2, alicki1977markov, roga2010davies}, that the dynamics in the eigenbasis of the Hamiltonian corresponds to a classical Liovillian rate equation which, does not couple to the off-diagonal terms in the density matrix. Here, the transition rates  $ P(k,m) = \sum_\alpha |S_{km}^\alpha|^2 G^\alpha(\epsilon_k - \epsilon_m)$ of the process correspond exactly to Fermi's golden rule for the perturbation operators $S^\alpha$, weighted with the bath correlation function \cite{alicki1977markov}. This rate equation is often referred to as the Pauli master equation and it can easily be seen that the classical generator is given by
\be\label{Paulimaster}
	L_\beta =  \sum_{km}P(k,m) \Big{(}\proj{m} - \ket{k}\bra{m}\Big{)}.
\ee 
The dynamics in the eigenbasis of the Hamiltonian corresponds exactly to the case block with $\nu = 0$. For this block, the matrix (\ref{junk}) can be seen to give rise to the Dirichlet form
\be
\bra{f} \hat{\cE}^0 \ket{f} = \frac{1}{2}\sum_{ml} P(m,l)\sigma_l (f_{ll} - f_{mm})^2,
\ee
which is exactly  the classical Dirichlet form of the Pauli master equation with generator $L_\beta$. Hence the $\nu = 0$ block corresponds to the evolution populations in the energy eigenbasis, whereas all the other blocks $\nu > 0$ describe the dynamics of the coherences \cite{cohen2006quantum} that are an energy difference $\nu$ apart. 

We furthermore have the following block decomposition for the variance. 

\begin{proposition}
Let $\ket{f} \equiv f\otimes \1 \ket{\Omega}$ and  $\sigma = Z^{-1} e^{-\beta H}$, then the variance is given by
\be\label{Var-split}
\Var_\sigma(f,f) = \bra{f} \hat{V}_\sigma \ket{f}  \Sp \mbox{where,} \Sp \hat{V}_\sigma =  \bigoplus_{\nu}  \hat{V}_\sigma^\nu
\ee
is block diagonal with the same blocks as Dirichlet Matrix $\hat{\cE}$. The respective blocks can be written as 
\be
\hat{V}_\sigma^0 = \sum_{km} \sigma_{k}\sigma_{m} \Big{(}\proj{\underline{m}} - \ket{\underline{m}}\bra{\underline{k}}\Big{)},
\ee
and for all $\nu \neq 0$,
\be
\hat{V}_\sigma^\nu =  \sum_{\un{m} \in \hat{\nu}}\sigma(m_1,m_2) \proj{\un{m}}.
\ee 
\end{proposition}

\proof{ This follows directly from the representation of $\Gamma_\sigma(f) = \sum_{nm} \sigma(n,m) \Pi_n f \Pi_m$. We write $\Var_\sigma(f,f) = \tr{\Gamma_\sigma(f^\dag)f} - \tr{\sigma f}^2$. Due to the matrix vector identity $\Var_\sigma(f,f) = \bra{f} \hat{V}_\sigma \ket{f}$, we have that
\be
\hat{V}_\sigma = \sum_{n,m} \sigma(n,m) \Pi_n\otimes \Pi_m^T - \sigma_n \sigma_m \Pi_n \otimes \1 \proj{\Omega}  \Pi_m \otimes \1. \vspace{-0.3cm}
\ee 
If we only focus on the terms $m=n$ in the first summand, we obtain $\hat{V}_\sigma^0$, whereas the remaining summands give rise to the diagonal matrices $\hat{V}_\sigma^\nu$. \qed} 

\vspace{0.3cm}
The fact that both $\hat{\cE}$ and $\hat{V}_\sigma$ are block diagonal in the same basis with the same blocks, simplifies the derivation of a lower bound for the spectral gap greatly. Note, that since each block is independent, we just have to choose the minimal support number in each block. For the particular block with $\nu = 0$, this problem reduces to the classical problem discussed in \cite{fill1991eigenvalue,diaconis1991geometric,levin2009markov}. Furthermore, under the assumption of primitivity we have that the only matrices which are rank deficient are $\hat{\cE}^0$ and $\hat{V}^0_\sigma$. The kernel of these matrices is given by $\ket{\Omega}$. We summarize the observations of this section in the following theorem. 

\begin{theorem}
Let $\cL_\beta$ denote a Davies Liovillian for the system described by the Hamiltonian $H=  \sum_{k} \epsilon_k \proj{k}$ and fixed point $\sigma = Z^{-1}\exp(-\beta H)$.  Furthermore, let the coupling to the bath be given by the operators $\{S^{\alpha}\}$,
then the spectral gap $\lambda$ is lower bounded by
\be
	\lambda = \min(\lambda_{cl},\lambda_{QM}),
\ee 
where we have defined $\lambda_{cl} = \tau(\hat{V}^0_\sigma,\hat{\cE}^0)^{-1}$, i.e. the gap of the associated Pauli master equation.  Furthermore, we define the gap that corresponds to the off diagonals as  $\lambda_{QM} = \min_{\nu > 0} \tau(\hat{V}^\nu_\sigma,\hat{\cE}^\nu)^{-1}$.
\end{theorem} 

\Proof{ This follows directly form the decomposition into positive definite blocks and application of the splitting lemma. It suffices to restrict to values $\nu \geq 0$, since 
the spectra of the matrices for $\nu >0 $ and $\nu < 0$ coincide.\qed}

\subsection{Does the gap $\lambda_{cl}$ of the classical Pauli master equation suffice?}

\begin{figure}[h]
\begin{center}
\resizebox{0.85\linewidth}{!}
{\includegraphics{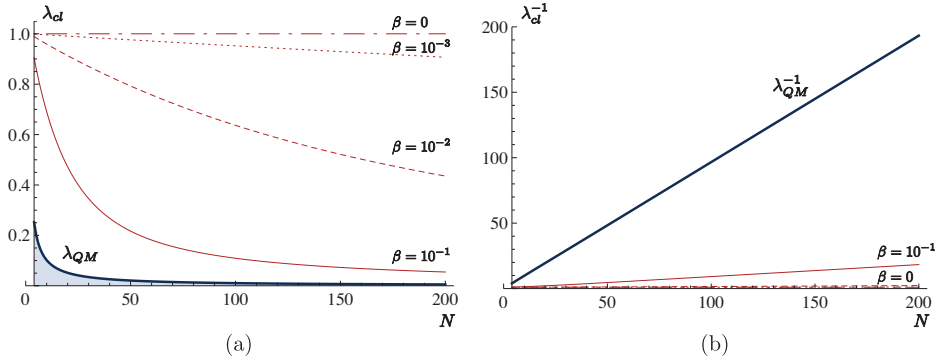}}
\caption{\label{fig:counter_example} We depict the eigenvalues $\lambda_{cl}$ and $\lambda_{QM}$ of the Davies generator which corresponds to the Hamiltonian $H$ and coupling operator $S$ as given in (\ref{def:counter_example}) for different values of $N = 4,\ldots,200$. The eigenvalues and their inverses were plotted for different values of $\beta = 0, 10^{-3},10^{-2},10^{-1}$. $\lambda_{QM}$ is independent of $\beta$ and always scales as $N^{-1}$, whereas the scaling of $\lambda_{cl}$ does indeed depend on the temperature. For illustration purposes we have chosen a particular  function $G(\omega) = (1 + e^{\beta \omega})^{-1}$, which is motivated from classical Glauber dynamics. The coupling $\gamma$ was adjusted to obtain for $\beta = 0$ the value $\lambda_{cl} = 1$. Note, that the quantum mechanical function $G(\omega)$ depends on the particular bath operators in general and generally differs from the one given above. A good study of different functions can be found in \cite{Breuer}}
\end{center}
\end{figure}

The steady state of the Davies generator can be found as an eigenvector in the block with the Bohr frequency $\nu = 0$. This block, as described previously,  corresponds to a classical master equation which describes the time evolution of the diagonal elements of the density matrix in the eigenbasis of the Hamiltonian. Since the steady state can be found in this block it is tempting to conjecture, that it is in fact always $\lambda_{cl}$  that determines the mixing time of the quantum mechanical Davies generator. Hence, if this conjecture were true, it would suffice to consider only the pair $(V^0,{\cal E}^0)$, to derive lower bounds to the spectral gap, and the classical tools would accomplish this task. However, here we provide a simple counter example and show that this can not hold in general and that the full dynamics have to be considered. Consider therefore the Hamiltonian given by
\be\label{def:counter_example}
H = \sum_{a=1}^N a \; \proj{a}, \Sp \mbox{ and the operator,}  \Sp  S = \frac{\gamma}{\sqrt{N}} \sum_{a,b=1}^N \ket{a}\bra{b},
\ee
which couples the system to the bath. Note, that the operator norm of $\| H \| = N$ scales extensively in the system size, whereas $S$ only scales as $\| S \| = \sqrt{N}$.  We will later provide a simple physically motivated example with the same scaling. Let us first consider the extremal case of $\beta = 0$. The eigenvalues of the steady state are given by $\sigma_n = N^{-1}$ and furthermore $G(\omega) = \mbox{const}  \equiv g$. Let us now compute the blocks corresponding to the first two Bohr frequencies, $\nu = 0,1$. We have that,
\bq
\hat{\cE}^0 &=&  \sum_{n,m=1}^N \frac{g\gamma^2}{N^2}\left(\proj{n} - \ket{n}\bra{m} \right)  =  g\gamma^2 \hat{V}^0\\ \hat{\cE}^1 &=&  \sum_{n=1}^{N-1}\frac{g\gamma^2}{N^2} \proj{n} -  \sum_{n,m=1}^{N-1} \frac{g\gamma^2}{N^2}\left(\proj{n} - \ket{n}\bra{m} \right) \Sp \mbox{and,} \; \hat{V}^1 =  \sum_{n=1}^{N-1} \frac{1}{N} \proj{n}.
\eq
The support can be calculated exactly by comparison. In this particular case, we obtain that 
\be
\lambda_{cl} = g\gamma^2, \Sp \mbox{whereas} \Sp \lambda_{QM} = g\frac{\gamma^2}{N}.
\ee
We have a clear separation between the two eigenvalues since the classical gap is constant, whereas the quantum gap decays linearly in the system size $N$.  Note, however, that this strong difference has occurred in the infinite temperature regime. When the temperature is finite, the separation between $\lambda_{cl}$ and $\lambda_{QM}$ depends on the particular form of the bath correlation function $G(\omega)$. In numerical experiments we found, that we always have that $\lambda_{cl} \geq \lambda_{QM}$, however, with the difference that the classical gap $\lambda_{cl}$ can also decay as $N^{-1}$ for low temperatures. The numerical findings are depicted in Fig.~\ref{fig:counter_example}.

\vspace{0.5cm}
\vspace{0.5cm}

\section{Lower bounds to the spectral Gap}\label{sec:lowerBounds}

The transition rates in the classical block define a transition graph between the different eigenstates. That is, we define a graph $(\{m\},E^0)$ between the eigenstates of the Hamiltonian $H$ by $(m,n) \in E^0$, if the corresponding transition element $P(m,n) = \sum_{\alpha}G^\alpha(\epsilon_m - \epsilon_n)|S^\alpha_{m,n}|^2 > 0$. Hence the matrix $\hat{\cE}^0$ can be interpreted as the weighted Laplacian matrix of this graph. Likewise the matrix $\hat{V}^0_\sigma$ is given by the Laplacian of the complete graph $K^0$. It is customary to define an embedding of the graph $E^0$ into the complete graph by defining for every pair of vertices $(a,b)$ a path $\gamma_{ab}$ that connects these two vertices and only traverses the links of $E^0$. The length $|\gamma_{ab}|$ of this path amounts to the number of edges in $E^0$ which are traversed by this path.\\

Let us turn to the matrix blocks with $\nu > 0$ as defined in (\ref{junk}). If we reacall the definition of $G^\alpha_{m_i}$, these blocks can be written as 
\bq
\hat{\cE}^\nu &= &  \sum_k \sum_{\un{m} \in \hat{\nu};\alpha} \frac{1}{2}\sum_{i=1}^2 G^\alpha(\epsilon_k - \epsilon_{m_i})  \sigma(m_1,m_2) |S^\alpha_{m_i,k}|^2  \; \;  \proj{\un{m}} \nonumber \\ & -& \sum_{\un{m}\un{l} \in \hat{\nu};\alpha} G^\alpha(\epsilon_{l_1} - \epsilon_{m_1})\sigma(m_1,m_2) \; \overline{S}^\alpha_{l_2, m_2} S^\alpha_{l_1,m_1} \; \ket{\un{l}}\bra{\un{m}}.
\eq
The sum over $k$ runs over all possible eigenstates of the full Hamiltonian $H$. The sum over the tuples $\underline{n} \in \hat{\nu}$ is more restrictive and therefore summands remain that can not be paired up as elements in $\hat{\nu}$. The remaining elements in the diagonal can be split off, and we write
\be
\hat{\cE}^\nu = \sum_{\un{m} \in \hat{\nu}} \phi^\nu_{\underline{m}} \; \;  \proj{\un{m}} + \sum_{ \un{m} \leq \un{l}} M^\nu_{\un{l}\un{m}},
\ee 
where we have defined
\be
\phi_{\underline{m}}^\nu = \frac{1}{2} \sum_{i=1}^2 \sum_{n_i \notin \hat{\nu}_i ;\alpha} G^\alpha(\epsilon_{n_i} - \epsilon_{m_i} ) |S_{m_i,n_i}^\alpha|^2 \sigma(m_1,m_2).
\ee
In the sum that defines $\phi_{\underline{m}}^\nu$ we have introduced the notation $n_i \notin \hat{\nu}_i$. By this we refer to the following: The set $\hat{\nu}_i$ corresponds to the energy labels that are in the $i$'th position of the tuples $(n_1,n_2) \in \hat{\nu}$, e.g for $\hat{\nu}_2$ we consider all the possible values of $n_2$ occurring in this set. These labels do not always enumerate all possible energy labels $\{k\}$. Hence the sum over $n_i \notin \hat{\nu}_i$ corresponds to the remaining energy labels that do not occur in $\hat{\nu}_i$. Moreover, we have defined the matrix 
\bq
M^\nu_{\un{l}\un{m}} = \sum_\alpha G^\alpha(\epsilon_{l_1} - \epsilon_{m_1})\sigma(m_1,m_2)\left[ \frac{1}{2} \left(|S^\alpha_{m_1,l_1}|^2 + |S^\alpha_{m_2,l_2}|^2 \right) \left(\proj{\underline{m}} + \proj{\underline{l}}\right)\right. \nonumber \\ \left . - \overline{S}^\alpha_{m_2,l_2}{S}^\alpha_{m_1,l_1} \ket{\underline{m}}\bra{\underline{l}} - \overline{S}^\alpha_{l_2,m_2}{S}^\alpha_{l_1,m_1} \ket{\underline{l}}\bra{\underline{m}} \right ].
\eq
This matrix is effectively two dimensional and can be diagonalized easily. One readily obtains the eigenvalues $\lambda_{\pm}(\underline{l},\underline{m})$, which correspond to the eigenvectors $2^{-1/2}\left(\ket{\underline{m}} - e^{i\theta_{ml}}\ket{l}\right)$ for $\lambda_{+}$ and $2^{-1/2}\left(\ket{\underline{m}} + e^{i\theta_{ml}}\ket{l}\right)$ for $\lambda_{-}$ respectively. The phases $e^{i\theta_{ml}}$ are chosen appropriately. 
\bq \label{eigenv_link}
\lambda_{\pm}(\underline{l},\underline{m}) =  \sum_{\alpha}G^\alpha(\epsilon_{l_1} - \epsilon_{m_1})\frac{1}{2}(|S^\alpha_{m_1,l_1}|^2 + |S^\alpha_{m_2,l_2}|^2 ) \sigma(m_1,m_2) \nonumber \\ \pm \left |\sum_\alpha G^\alpha(\epsilon_{l_1} - \epsilon_{m_1})\overline{S}^\alpha_{m_2,l_2}S^\alpha_{m_1,l_1} \right| \sigma(m_1,m_2).  
\eq
Note, that both eigenvalues $\lambda_{\pm}(\underline{l},\underline{m})$ are always positive due to the triangle inequality. If the product of matrix elements $\overline{S}^\alpha_{m_2,l_2}S^\alpha_{m_1,l_1} $ is always real and positive, or if only a single $\alpha$ contributes to the full sum, the eigenvalues can be simplified to
\be
\lambda_{\pm}(\underline{l},\underline{m}) =  \sum_{\alpha}G^\alpha(\epsilon_{l_1} - \epsilon_{m_1})\sigma(m_1,m_2)\frac{1}{2}(|S^\alpha_{m_1,l_1}| \pm |S^\alpha_{m_2,l_2}|)^2. 	
\ee
If we consider the case $\nu = 0$ for a moment, i.e. $m_1 = m_2$ and $l_1 = l_2$, we see that the eigenvalues reduce to $\lambda_{-}(l,m) = 0$ whereas $\lambda_{+}(l,m) = P(l,m)\sigma_m$. For this special case all the $\phi_{\underline{m}}^\nu$ vanish and we are again left with the classical rate equation.

Note, that the matrices $\hat{V}_\sigma^\nu$ are very different for  $\nu = 0$ and $\nu > 0$. In the former case the matrix corresponds to the weighted Laplacian of the complete graph, whereas for $\nu > 0$ the matrix is simply diagonal. We therefore have to consider these two cases separately. We first start with a simple bound for the spectral gap, which is obtained by considering first the canonical path lemma \cite{fill1991eigenvalue,diaconis1991geometric} for the classical block and then applying a Gershgorin bound \cite{Bhatia} for the remaining blocks. These bounds turn out to be quite useful already and very easy to apply. We will later turn to more complicated bounds, which have the advantage of being more robust when the simpler bounds fail. That occours for example, when the simple bounds predict a vanishing spectral gap even though the map is primitive. 

\begin{theorem}\label{MainBound} Let $\cL_\beta$ denote a Davies Liovillian for the system described by the non-degenerate  Hamiltonian $H=  \sum_{k} \epsilon_k \proj{k}$, then the spectral gap $\lambda$ of $\cL_\beta$ is lower bounded by
\be
	\lambda \geq \min\left(\frac{1}{\tau^0}\; ,\; \Lambda_{QM}\right),
\ee 
where the two constants are defined as follows.
\begin{itemize}
\item The gap in the block $\nu = 0$ can be bounded by
\be
\tau^0 = \max_{(n,m) \in E^0} \frac{1}{P(n,m)\sigma_m} \sum_{\gamma_{ab} \ni (n,m)}\sigma_a \sigma_b |\gamma_{ab}|, 
\ee
with $P(n,m) = \sum_\alpha G^\alpha(\epsilon_n - \epsilon_m)|S_{n,m}^\alpha|^2$.

\item
Furthermore $\Lambda_{QM}$ is obtained from the optimization 
\bq
	\Lambda_{QM} = \min_{(m_1 < m_2)} \frac{1}{2} \Lambda_{\underline{m}},
\eq
where,
\bq \label{LambdaOmg}
\Lambda_{\underline{m}} =  \sum_{\un{n} \in \hat{\nu};\alpha}G^{\alpha}(\epsilon_{n_1} - \epsilon_{m_1} ) \left( |S^\alpha_{n_1m_1}| - |S^\alpha_{n_2m_2}| \right)^2  +\sum_{i=1}^2 \sum_{n_i \notin \hat{\nu}_i;\alpha} G^\alpha(\epsilon_{n_i} - \epsilon_{m_i} )  |S^\alpha_{n_im_i}|^2, 
\eq
with $\nu = \epsilon_{m_2} - \epsilon_{m_1}$ and the set $\hat{\nu}$ corresponds to all pairs $(n_1,n_2)$ that give rise to the same energy difference.
\end{itemize}
\end{theorem} 

\proof{ The matrix pencil $(\hat{V}^0_\sigma,\hat{\cE}^0)$ just reduces to the well known classical problem, for which good bounds are already known \cite{fill1991eigenvalue,diaconis1991geometric}. One particular bound is given by the canonical path lemma . The proof follows from a clever application of the Cauchy-Schwartz inequality and we will defer the reader to the reference \cite{fill1991eigenvalue}.

The bound on the support number for $\nu > 0$, and by that on $\lambda_{QM}$ is a direct consequence of Gershgorin's theorem \cite{Bhatia}. Note, that we have
\bq
\tau^\nu \hat{\cE}^\nu  - \hat{V}^\nu_\sigma =  \sum_{\un{m} \in \hat{\nu}} \tau^\nu \phi^\nu_{\underline{m}}  - \sigma(m_1,m_2) \; \;  \proj{\un{m}} + \sum_{(\un{m},\un{l}) \in E^\nu} \tau^\nu M^\nu_{\un{m}\un{l}}.	
\eq
We can now apply Gershgorin's theorem and see, that the eigenvalues of $\tau^\nu \hat{\cE}^\nu  - \hat{V}^\nu_\sigma$, have to lie within the intervals $[g^{\underline{m}}_{+},g^{\underline{m}}_{-}]$ determined by 
\bq
g^{\underline{m}}_{\pm} = \left(\tau^\nu (\phi^\nu_{\underline{m}} + \sum_{\underline{n} \in \hat{\nu}} \lambda_{\pm}(\underline{m},\underline{n})) - \sigma(m_1,m_2)\right).
\eq 
If we choose for each $\underline{m}$ the lower bound $ g^{\underline{m}}_{-}$ and observe, that we always have 
\be
\lambda_{-}(\underline{m},\underline{n})) \geq \sum_\alpha G^{\alpha}(\epsilon_{n_1} - \epsilon_{m_1} ) \left( |S^\alpha_{n_1m_1}| - |S^\alpha_{n_2m_2}| \right)^2\sigma(m_1,m_2),
\ee
we are left with the constrained $\tau^\nu\frac{1}{2}\Lambda_{\underline{m}} - 1 \geq 0$. If we now choose $\tau^\nu \geq \Lambda_{QM}^{-1}$ as defined in the lemma, we are ensured that the difference between the two matrices is positive semi-definite and we are left with the lower bound as stated in the lemma. \qed}

\vspace{0.3cm}
The Gershgorin bound on the blocks for $\nu > 0$ is exact when  $|\hat{\nu}| = 1$. This occurs, when the Bohr frequencies are not degenerate and the energies are unevenly spaced. In general one can expect that this bound provides a good estimate, when the Bohr frequencies are not strongly degenerate and when the diagonals are strictly dominant.

As we have mentioned before, c.f. lemma \ref{lem:W-bound}, a good method for finding bounds on the support number can be obtained from a suitable factorization of the matrices $\hat{\cE}^\nu$ and $\hat{V}^\nu_\sigma$. We now proceed to derive more robust bounds to the spectral gap that also yield satisfactory answers when the aforementioned  bound fails.  First, we focus on the classical block and show how a a different bound on the classical gap can be derived. This was shown already in \cite{boman2003support}, and we include the proof here only for completeness. Moreover, other well known  spectral gap bounds  can be derived this way, as for instance, the congestion dilation lemma \cite{spielman2003solving,chen2005obtaining}. We then proceed  to derive a new bound for the off diagonal blocks  with $\nu > 0$.

\begin{lemma}\label{lem-classicalDec}
For $\hat{\cE}^0$ and $\hat{V}_\sigma^0$, we have the following decomposition into $\hat{\cE}^0 = A^0 {A^0}^\dag$ and  $\hat{V}_\sigma^0 = U^0 {U^0}^\dag$, with $W^0$ so that $A^0 W^0 =  U^0$. We have defined $P(i,j) = \sum_\alpha G^\alpha(\epsilon_i - \epsilon_j)|S^\alpha_{i,j}|^2$ so that
\bq
A^0   &=& \sum_{i < j } \sqrt{P(i,j)\sigma_j}\left(\ket{i}  - \ket{j}\right)\bra{ij}, \Sp \mbox{and} \Sp
U^0  = \sum_{i < j } \sqrt{\sigma_i\sigma_j}\left(\ket{i} -  \ket{j}\right)\bra{ij},  \\
W^0 &=& \sum_{i < j } \ket{w_{ij}}\bra{ij}.  \Sp \mbox{with} \Sp
\ket{w_{ij}} = \sum_{(a,b) \in \gamma_{ij}} \sqrt{\frac{\sigma_i\sigma_j}{P(a,b)\sigma_b}}\; \mbox{sign}(b-a)\ket{(a,b)_{>}}.
\eq
Here, $\gamma_{ij}$ denotes a path in $E^0$ connecting the vertices labeled by $i,j$ through all the links $(a,b) \in E^0$ and $(a,b)_{>}$ corresponds to an ordering of the pair $(a,b)$.
\end{lemma}

\proof{ We can write
\bq
\hat{\cE}^0  =  \sum_{i < j} P(i,j)\sigma_j \left(\ket{i}  - \ket{j}\right)\left(\bra{i}  - \bra{j}\right) \Sp \mbox{and} \Sp
\hat{V}^0_\sigma = \sum_{i< j} \sigma_i\sigma_j \left(\ket{i}  - \ket{j}\right)\left(\bra{i}  - \bra{j}\right).
\eq 
We associate to each edge $(i,j)$ in the graph $E^0$ a projector onto the vector $\sqrt{P(i,j)\sigma_j}\left(\ket{i}  - \ket{j}\right)$.  For each edge we choose a state $\ket{ij}$ in a larger (edge) space. So in the decomposition with $A^0$ and $U^0$ every row corresponds to a vertex index, whereas every column corresponds to a link.  It is therefore easy to verify that $\hat{\cE}^0 = A^0 {A^0}^\dag$ and $\hat{V}_\sigma^0 = U^0 {U^0}^\dag$. The matrix $W^0$ can  be seen as an embedding from the graph $A^0$ into the complete graph $U^0$. It is easily verified that $A^0 W^0 = U^0$, since the positive and negative contributions in $\sqrt{P(i,j)\sigma_j}\left(\ket{i}  - \ket{j}\right)$ cancel appropriately along the path. \qed }

\vspace{0.3cm}
The blocks behave differently for $\nu > 0$, since we have to find a factorization for two full rank matrices.We have seen earlier, that he matrices $\hat{\cE}^\nu$ are all diagonal dominant and of course positive definite.  A suitable factorization can be found by first splitting off the remaining summands on the diagonals $\phi^\nu_{\underline{m}}$ and then considering the individual matrices $M_{\underline{l},\underline{m}}^\nu$.\\

In order to construct this decomposition, we need to introduce some new notation. 

\begin{figure}[h]
\begin{center}
\resizebox{0.3\linewidth}{!}
{\includegraphics{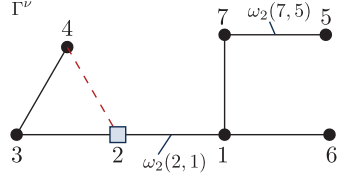}}
\caption{\label{fig:tree} The figure depicts the construction of the tree $T^\nu$ from the graph $E^\nu$. The tree in this example is obtained by removing the dashed link labeled by $(2,4)$. The two weights shown in the figure can be computed from the tree structure by $\omega_2(7,5) = \phi_5 + \lambda_{-}(7,5)$ and $\omega_2(2,1) = \lambda_{-}(2,1) + \phi_1 + \phi_5 + \phi_6 + \phi_7 + 2\left(\lambda_{-}(7,5) +\lambda_{-}(1,7) + \lambda_{-}(1,6)\right)$ This sum is uniquely determined by the tree $T^\nu$, and corresponds to summing up all $\phi_n$ that live on the branches lower than the current node measured by the marked vertex (2 in this example). This sum is denoted by ${fill}^m_{n,l}(T^\nu)$ for a link $(n,l)$, with the marked vertex $m$.}
\end{center}
\end{figure}

We now associate to each block a graph $(\hat{\nu},E^\nu)$, where two vertices, i.e. elements in $\hat{\nu}$ are connected when the ${+}$ eigenvalue, defined in (\ref{eigenv_link})  is different from zero. This means, that $\underline{n},\underline{m} \in \hat{\nu}$ are connected when $\lambda_{+}(\underline{n},\underline{m}) > 0$.  For $\nu = 0$, this just corresponds to the previously introduced graph $E^0$. For $\nu > 0$, it corresponds to the graph that is induced by the $\hat{\cE}^\nu$ when taken as a weighted adjacency matrix. Furthermore, we need to introduce an associated tree $T^\nu$, which can be obtained from the graph $E^\nu$ by deleting edges to break up any closed cycle. For every cycle in $E^\nu$, we will only need to delete a single edge. This construction is not unique, and differently constructed trees may give rise to different lower bounds on the  gap.  A graphical representation can be found in Fig.~\ref{fig:tree}. 

\begin{lemma}\label{lem-off-factorization}
Let $\hat{\cE}^{\nu}$ denote the $\nu$'th block of the Dirichtet matrix, and $\hat{V}^\nu_\sigma$ the corresponding block in the matrix associated with the variance. A particular decomposition into matrices $ \hat{\cE}^{\nu} = {A^\nu}{A^\nu}^\dag$  and $\hat{V}^\nu_\sigma = U^\nu{U^\nu}^\dag$ with ${A^\nu}W^\nu  = U^\nu$ is given by
\bq
A^\nu =  \sum_{\underline{m} \in \hat{\nu}} \sqrt{\phi^\nu_{\underline{m}}}\proj{\underline{m}} + \frac{1}{2}\sum_{ (\underline{m},\underline{l}) \in E^\nu} 
&& \sqrt{\lambda_+(\underline{l},\underline{m})} \left(\ket{\underline{m}} - e^{i\theta_{m,l}}\ket{\underline{l}}\right)\bra{(\underline{m},\underline{l})^+} \nonumber \\
&+& \sqrt{\lambda_-(\underline{l},\underline{m})} \left(\ket{\underline{m}} + e^{i\theta_{m,l}}\ket{\underline{l}}\right)\bra{(\underline{m},\underline{l})^-}
\eq
\bq
U^\nu &=& \sum_{\underline{m} \in \hat{\nu}} \sqrt{\sigma(m_1,m_2)}\proj{\underline{m}} \Sp \mbox{and} \Sp 
W^\nu =  \sum_{\underline{m} \in \hat{\nu}} \frac{\sqrt{\sigma(m_1,m_2)}}{2 N_\nu} \ket{w_{\underline{m}}}\bra{\underline{m}},
\eq
where we require that the vectors $\{\ket{\underline{m}},\ket{(\underline{m},\underline{l})^+},\ket{(\underline{m},\underline{l})^-} \}$ are mutually orthonormal. The normalizing constant is given by
\be
N_{\nu} =  \sum_{\underline{l} \in \hat{\nu} } \phi^\nu_{\underline{l}} + \sum_{(\underline{l},\underline{n}) \in T^\nu} \lambda_{-}(\underline{l},\underline{n}). 
\ee
The vectors $\ket{w_{\underline{m}}}$  which define the matrix $W^\nu$ are defined as follows: Consider a tree $T^\nu$, which is obtained from the graph $E^\nu$ by deleting a link in every closed loop, then  
\bq\label{evelVec}
\ket{w_{\underline{m}}} = \sum_{\underline{l} \in \hat{\nu}} \sqrt{\phi^\nu_{\underline{l}}} e^{-i\hat{\theta}_{ml}}\ket{\underline{l}} 
+ \sum_{(\underline{l},\underline{n}) \in T^\nu} e^{-i\hat{\theta}_{nl}}\left( \sqrt{\lambda_{-}(\underline{l},\underline{n})} \ket{(\underline{l},\underline{n})^-} +\frac{\omega^\nu_{\underline{m}}(\underline{l},\underline{n}) }{\sqrt{\lambda_{+}(n,l)}} \ket{(\underline{l},\underline{n})^+}\right), \nonumber
\eq
with
\be\label{weightOmega}
\omega^\nu_{\underline{m}}(\underline{l},\underline{n})  = \lambda_{-}(\underline{l},\underline{n}) +  \phi^\nu_{\underline{n}} + \sum_{{fill}^m_{n,l}(T^\nu)} 
\phi_{\underline{b}}^\nu +  2 \lambda_{-}(\underline{a},\underline{b}).
\ee
\end{lemma}  

\vspace{0.3cm}
Before we proceed to prove the lemma, let us briefly explain the notation. In particular the index ${fill}^m_{n,l}$ of the summation for the tree $T^\nu$ needs explanation. The tree $T^\nu$ is obtained from the graph $E^\nu$ by removing edges that complete a cycle. As was stated earlier, this construction is of course not unique. However, once the tree is constructed, the summation over ${fill}^m_{n,l}(T^\nu)$ is uniquely defined. The construction of some weight  $\omega^\nu_{\underline{m}}(\underline{l},\underline{n})$  at the link $(\underline{l},\underline{n}) \in T^\nu$  with a fixed reference node $\underline{m}$ can also be understood recursively. We write 
\be\label{recursiveOmega}
\omega^\nu_{\underline{m}}(\underline{l},\underline{n}) = \phi^\nu_{\underline{n}} + \lambda_{-}(\underline{l},\underline{n}) + \sum_{ \underline{r} \sim \underline{n}} \left( \omega^\nu_{\underline{m}}(\underline{n},\underline{r})  +   \lambda_{-}(\underline{n},\underline{r}) \right). 
\ee
Here we sum only over links $(\underline{n},\underline{r})$ that are directly connected to the node $\underline{n}$ and are not the link $(\underline{l},\underline{n})$ itself. We have assumed that to get to vertex $\underline{m}$ from vertex $\underline{n}$, we need to traverse the link $(\underline{l},\underline{n})$. If one therefore carries out the summation explicitly, we have to sum over all the branches of the tree that lead up to the link $(\underline{l},\underline{n})$ which we have to cross to reach the vertex $\underline{m}$. We sum for every vertex on these branches the corresponding weight $\phi^\nu_{\underline{b}}$ and for every link that leads up to $(\underline{l},\underline{n})$ the weight $2\lambda_{-}(\underline{a},\underline{b})$. The full summation therefore corresponds to eqn. (\ref{weightOmega}), where we picture that the summation fills up the remaining branches that lead up to the link $(\underline{l},\underline{n})$. The construction of the tree as well as the summation for some weights is explained for an example graph in Fig.~\ref{fig:tree}. \\

\vspace{1cm}
\proof{ 
Given the eigenvalue decomposition of the small two-dimensional matrices $M^\nu_{\underline{l},\underline{m}}$, it can be verified by direct multiplication that the matrix $A^\nu$ is a valid root of the Dirichlet block $\hat{\cE}^\nu$, since we have defined the the vectors $\{\ket{\underline{m}} , \ket{(\underline{m},\underline{l})^+}, \ket{(\underline{m},\underline{l})^-}\}$ to be orthonormal. The factorization of $\hat{V}^\nu$ is trivial.  The structure of the matrix $\hat{V}^\nu$ for $\nu > 0$ is quite different from the case when $\nu =0$ since it is a mere diagonal matrix. The vectors $\ket{w_{\underline{m}}}$ therefore have to be chosen as the appropriately normalized dual vectors to the row vectors of $A^\nu$. A direct calculation shows that
\bq
\bra{\underline{l}}A^\nu\ket{w_{\underline{m}}} = \frac{\sqrt{\sigma(m_1,m_2)}}{2N^\nu}\left( \phi^\nu_{\underline{l}} + \mbox{sign}(\underline{l},\underline{m})\left(\omega^\nu_{\underline{m}}(\underline{l},\underline{m}) + \lambda_{-}(\underline{l},\underline{m})\right) \right. \nonumber \\ + \sum_{ \underline{r} \sim \underline{n}} \left. \left( \omega^\nu_{\underline{m}}(\underline{n},\underline{r})  +   \lambda_{-}(\underline{n},\underline{r}) \right) \right),
\eq
where $\mbox{sign}(\underline{l},\underline{m}) = 1$, when $\underline{m} = \underline{l}$ and $\mbox{sign}(\underline{l},\underline{m}) = -1$ otherwise. This can always be achieved by fixing a certain ordering in the tree and choosing the phases $\hat{\theta}_{ml}$ appropriately. Since we are considering the summation over a tree and do not have to worry about any closed loops, these phases can be assigned uniquely. We see that with the recursive definition of $\omega^\nu_{\underline{m}}(\underline{l},\underline{m})$, as given in eqn. (\ref{recursiveOmega}), we have $\bra{\underline{l}}A^\nu\ket{w_{\underline{m}}} = 0$ whenever $l$ and $m$ differ.For the particular case when they are the same we immediately have that $\bra{\underline{m}}A^\nu\ket{w_{\underline{m}}} = \sqrt{\sigma(m_1,m_2)}$.  \qed }

\vspace{0.3cm}  
The lemmata \ref{lem-classicalDec} and \ref{lem-off-factorization} can now immediately be used to derive upper bounds to the support number $\tau$ by making use of the norm bounds on $W^\nu$. The bound presented here is only one possible bound, that can be obtained from the factorization and is admittedly more complicated than the Gershgorin bound of theorem \ref{MainBound}. However, these bounds seem to be tighter in many cases and moreover provide satisfactory answers even when the matrices $\hat{\cE}^\nu$ are not strictly diagonal dominant in every row. 

\begin{theorem}\label{robustBound}
Let $\cL_\beta$ denote a Davies Liovillian for the system described by the non-degenerate  
Hamiltonian $H=  \sum_{k} \epsilon_k \proj{k}$, then the spectral gap $\lambda$ of $\cL_\beta$ is lower bounded by
\be
	\lambda \geq \min\left(\frac{1}{\hat{\tau}^0} ,\; \hat{\Lambda}_{QM}\right),
\ee 
where the two constants are defined as follows:
\begin{itemize}
\item We have for $\hat{\tau}^0$ the following congestion dilation bound
\be\label{TauHat}
\hat{\tau}^0 = \left(\max_{(a,b) \in K^0} \sum_{(n,m) \in \gamma_{ab}} \sqrt{\frac{\sigma_a\sigma_b}{P(n,m)\sigma_m}} \right)\left(\max_{(a,b) \in E^0} \sum_{\gamma_{nm} \ni (a,b)} \sqrt{\frac{\sigma_n\sigma_m}{P(a,b)\sigma_b}}\right), 
\ee
where $K^0$ denotes the complete graph on the all the energy eigenstates of $H$.

\item
Furthermore,  we obtain $\hat{\Lambda}_{QM}$ from choosing the minimal  
\bq
	\hat{\Lambda}_{QM} = \min_{ \nu }  \hat{\Lambda}_\nu,
\eq
\vspace{-0.3cm}with
\bq \label{LambdaNu}
\hat{\Lambda}_\nu^{-1} = \sum_{\underline{m} \in \hat{\nu}}\frac{\sigma(m_1,m_2)}{4N_\nu^2}\left(\sum_{\underline{l} \in \hat{\nu}}\phi^\nu_{\underline{l}} + 
\sum_{(\underline{l},\underline{n}) \in T^\nu} \left(\lambda_{-}(\underline{l},\underline{n}) + \frac{\left[\omega^\nu_{\underline{m}}(\underline{l},\underline{n})\right]^2 }{\lambda_{+}(\underline{l},\underline{n})}\right) \right).
\eq
\end{itemize}
The notation was introduced in the proceeding paragraph.
\end{theorem}

\proof{ These bounds follow directly from the decomposition of the matrix pencils $(\hat{V}^0_\sigma,\hat{\cE}^0)$ and $(\hat{V}^\nu_\sigma,\hat{\cE}^\nu)$ as given in lemma \ref{lem-classicalDec} and \ref{lem-off-factorization} respectively. To obtain the bound (\ref{TauHat}), proved in \cite{boman2003support}, we apply the upper bound $\|W^0\|^2_{2\raw2} \leq \|W^0\|_{1}\|W^0\|_{\infty}$ \cite{Bhatia} and observe that the column space corresponds to the edge space of the graph $E^0$ whereas the row space corresponds to that of $ \hat{V}^0_\sigma $  and thus to the complete graph $K^0$. The bound on $\hat{\Lambda}_{QM}$ is obtained by upper bounding the $2\raw2$ norm of $W^\nu$ by the Frobenius norm. \qed} 

The bounds provided in theorem \ref{robustBound}  are only two possible bounds that can be obtained from the matrix decomposition we have provided. In fact other bounds on the $2\raw2$ norm are known \cite{chen2005obtaining}, and in some cases it may be preferable to use these as opposed to the ones provided.

%
% Applications
%
\section{Applications and Example systems}\label{sec:Applicaitons}
Before we proceed to discussing different examples for which the spectral gap $\lambda$ can be bounded, we need to point out a property of the bath function $G(\nu)$. This function is generally determined by the bath the system couples to. A detailed account can be found in \cite{Breuer}. We  will, however, assume for illustration purposes that the function is always given by 
\be
	G(\omega) = \frac{1}{1 +  e^{\beta \omega}}.
\ee
The motivation stems from the function of classical Glauber dynamics. An actual physical bath correlation function may indeed be very different. However, this function already satisfies some important properties. The function is always upper bounded by $G(\omega) \leq 1$ and positive $0 \leq G(\omega)$. However, it does not posses a lower bound since for all finite $\beta \geq 0$ we have that $\lim_{\omega \raw \infty} G(\omega) = 0$. 
 
\subsubsection{Example:  Truncated harmonic oscillator }
Since our bounds apply to finite state spaces, let us consider a harmonic oscillator on a truncated Hilbert space $\bC^{D+1}$. That is, we only consider the first $D$ eigenmodes of the oscillator. Moreover, we couple to the bath only by the position of the particle, i.e. we assume that $S \propto \hat{x}$. We have that: 

\begin{itemize}
\item The Hamiltonian is given by: 
\be H = \epsilon a^\dag a = \sum_{n=0}^D \epsilon  n \; \proj{n} \ee

\item The system couples to the bath via:
\bq 
&& S = \gamma (a + a^\dag),  \Sp \mbox{where we have} \\
&& a^\dag = \sum_{n=0}^D \sqrt{n} \ket{n}\bra{n-1}. \eq
\end{itemize}
The systems steady state is given by
$ \sigma = Z^{-1} \sum_{n=0}^{D} e^{-Kn} \proj{n} $ with$ Z = (1 - e^{-K(D+1)})(1 - e^{-K})^{-1} $. The dynamics of the semi group are related to the matrix elements $S_{ab}$ of the coupling operators. 
We can compute the matrix elements in the Hamiltonian's eigenbasis as 
\be
	|S_{ab}|  = |\gamma|\left( \sqrt{a} \delta_{a-1,b} + \sqrt{b} \delta_{a,b-1}\right),
\ee
and naturally also 
\be
	|S_{ab}|^2 = \gamma^2\left( a \; \delta_{a-1,b} + b \; \delta_{a,b-1}\right).
\ee
This immediately induces the following coupling graph of the transition process:

\begin{figure}[h]
\begin{center}
\resizebox{0.815\linewidth}{!}
{\includegraphics{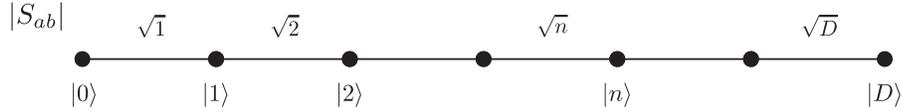}}
\caption{\label{fig:harm_osc} Interaction graph for the truncated harmonic oczilator. We consider a finite state space which is truncated at dimension $D$. The graph depicted corresponds to $E^0$. All other graphs, i.e. $E^\nu$, can obtained from this one.}
\end{center}
\end{figure}

Let us first look at the constant $\Lambda_{QM}$. For this energy level structure, it is simpler to enumerate all possible Bohr frequencies rather than comparing $\Lambda_{(n,m)}$ for different energy pairs.  We naturally have that $\nu \in \left\{0,1,\ldots,D\right\}$ and we can write:
\be
	\hat{\nu} = \left\{(n,n+\nu) | n = 0,\ldots,D-\nu \right\} \Sp \mbox{so that} \Sp |\hat{\nu}| = D-\nu + 1.
\ee
Now we can evaluate the constant $\Lambda^\nu_{(m,m+\nu)}$ as defined in (\ref{LambdaOmg}). Note, that the different sums read as $\un{n} \in \hat{\nu} \raw  n \in \{0,\ldots,D-\nu\}$ and furthermore $n_2 \raw \hat{\nu}_2 = n_2 \in \{0,\ldots,\nu\}$ as well as $n_1 \notin \hat{\nu}_1\raw n_1 \in \{D-\nu +1,\ldots,D\}$. The evaluation thus yields
\bq
\Lambda^\nu_m = \frac{\gamma^2}{2} \left\{ \begin{array}{ll} 
m=0 &:\; G(1)\left(1 - \sqrt{1 + \nu}\right) + G(-1)\nu\\
m \notin \partial\hat{\nu} &:\; G(1)\left(\sqrt{m + 1} - \sqrt{m + 1 + \nu} \right)^2 
+ G(-1)\left(\sqrt{m} - \sqrt{m + \nu} \right)^2  \\
m=D-\nu &:\; G(1)\left(D - \nu -1\right) + G(-1)\left(\sqrt{D} - \sqrt{D + \nu} \right)^2 
\end{array} \right.
\eq
where $G(1) = (1 + e^K)^{-1}$ as well as $G(-1) = (1 + e^{-K})^{-1}$. Recall that for $\Lambda_{QM}$ we have to consider $\nu > 0$. Looking at the equations we see that they become smaller the smaller $\nu$ is. Hence, we choose $\nu = 1$. It is easy to see that the difference in the square becomes smaller for larger $m$ as the difference between expressions vanishes. However this is only true for $m \neq D-\nu$. Hence, in the end we are left with 
\bq
\Lambda_{QM} &=& \frac{\gamma^2}{2} \left(G(1)\left(\sqrt{D-1} - \sqrt{D} \right)^2 
+ G(-1)\left(\sqrt{D-2} - \sqrt{D-1} \right)^2\right) \\
\Lambda_{QM} &\approx& \frac{\gamma^2}{8} \frac{1}{D}.
\eq
Let us focus on the classical gap $\lambda_{cl}$. For the classical transition rates we have 
\be
	P(n+1,n) = \frac{\gamma^2(n+1)}{1 + e^K}, 
\ee
which has to be paired with $\sigma_n = Z^{-1}e^{-Kn}$. Thus the bound for $\tau^0$ is given by
\bq
	\tau^0 &=& \max_{n} \frac{1}{P(n+1,n)\sigma_n} \sum_{\gamma_{ab} \ni (n,n+1)} \sigma_a \sigma_b |\gamma_{ab}|,\\
	&=&  \frac{1+e^K}{Z \gamma^2} \max_{n} \frac{e^{Kn}}{n+1} \sum_{a=0}^{n}\sum_{b=n+1}^D (b-a) e^{-K(a+b)}.
\eq
Direct evaluation and the application of some inequalities yields the following bound
\be
	\tau^0 \leq \frac{2D}{\gamma^2}\frac{1 + e^K}{e^K}.
\ee
Hence, in the end we are left with the bound 
\be
\lambda_{cl} \geq \frac{1}{1+e^{-K}} \frac{\gamma^2}{2} \frac{1}{D}.
\ee 
This leads to the total bound on the gap, which scales as $\lambda \geq O(\gamma^2D^{-1})$. This bound indeed agrees with the numerical experiments conducted where the same scaling was observed. We therefore see that the lower bound to the spectral gap does not depend on the system's temperature and we can find together with the previous result the following bound on the equilibration time $t_{mix} \leq {\cal O}(K\gamma^{-2}D^2\log(\epsilon^{-2}))$.% Note, that we have obtained this model from the truncation of a model on $\bL_2(\bR,d\mu)$. Hence, we expect therefore that the actual generator is gapless as we can see that the lower bound decays as $D \raw \infty$. 

\subsubsection{Example:  Particle on a line}
As the second example, let us consider a Fermion hopping on a line. The Hamiltonian, we consider stems from restricting the multiparticle problem $H = \sum_{r=1}^{N-1} (a_{r+1}^\dag a_{r} + h.c.) - g\sum_{r=0}^N a^\dag_r a_r$  with the coupling operator $S_k = \gamma a^\dag_k a_k$ to the single particle subspace. This ensures that the energy spectrum is non-degenerate. The system is described by
\begin{itemize}
\item
	the Hamiltonian
	\bq H = \sum_{r=1}^{N-1} \ket{r+1}\bra{r} -g\proj{r} + \ket{r}\bra{r-1} 
	  	 =  \sum_{k=1}^{N} \epsilon_k \proj{k}, \eq
	where the spectrum and the eigenvectors are given by
	
	\bq \epsilon_k &=& 2  \cos\left(\frac{k \pi}{N+1}\right) -g , \Sp \mbox{and} \\
	       \ket{k}        &=&       \sum_{n=1}^{N} \sqrt{\frac{2}{N+1}}  \sin \left(\frac{\pi}{N+1} k\; n\right)\ket{n}.\eq
\item 
	  The couplings to the environment in the single particle subspace are simply given by
	  \be S_n = \gamma \proj{n}. \ee
	   which in the eigenbasis corresponds to an all to all coupling in the interaction graph
	   \be S_n = \frac{2}{N+1} \sum_{pq=1}^N \sin \left(\frac{\pi}{N+1} p\; n\right) \sin \left(\frac{\pi}{N+1} q\; n\right) \ket{p}\bra{q}.\ee
	   We assume that $G_n(\omega) = G(\omega)$ is uniform for each coupling. 
\end{itemize}

\begin{figure}[h]
\begin{center}
\resizebox{0.75\linewidth}{!}
{\includegraphics{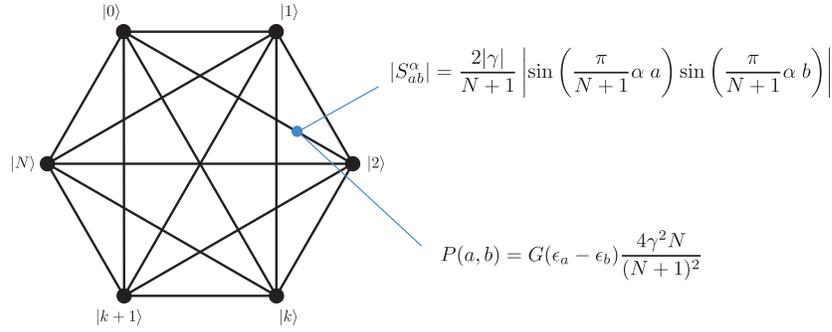}}
\caption{\label{fig:single_particle} Interaction graph for a single particle hopping on a line. The transitions are induced by the local densities which become very non local in the energy eigenbasis. Since the energies are very unevenly spaced, it is only the graph $E^0$ that is relevant.}
\end{center}
\end{figure}

Let us now focus on both, the classical as well as quantum transition amplitudes. We have that by substituting the label $\alpha$ for $n$, the following rates
\be
|S^\alpha_{ab}| = \frac{2|\gamma|}{N+1} \left| \sin \left(\frac{\pi}{N+1} \alpha \; a \right) \sin \left(\frac{\pi}{N+1} \alpha \; b \right)\right|
\ee
and furthermore 
\be
|S^\alpha_{ab}|^2 = \frac{4\gamma^2}{(N+1)^2} \sin^2 \left(\frac{\pi}{N+1} \alpha \; a \right) \sin^2\left(\frac{\pi}{N+1} \alpha \; b \right),
\ee 
which leads to the following classical transition probabilities
\bq
P(a,b) =  G(\epsilon_a - \epsilon_b)\sum_{\alpha=1}^{N} |S^\alpha_{ab}|^2 = G(\epsilon_a - \epsilon_b) \frac{4\gamma^2 N}{(N+1)^2}.
\eq
This leads to the coupling graph (Fig.~\ref{fig:single_particle}).

Let us consider the Bohr frequencies $\nu = \epsilon_a - \epsilon_b$. One can see, due to the transcendental nature of the differences between the cosines, each energy pair difference gives rise to a different Bohr frequency. This is very distinct from the model of the Harmonic oscillator, where the energy differences are evenly spaced. This leads to the fact, that all the matrices $\hat{\cE}^{\nu}$ are rank one and the Gershgorin bound on $\Lambda_{QM}$ becomes exact.\\

We therefore consider $\Lambda_{QM}$ first. A direct evaluation shows that 
\bq
	\Lambda_{(a,b)} &=& \frac{G(\epsilon_a - \epsilon_b)}{2}\sum_{\alpha=1}^N\left( |\sin \left(\frac{\pi}{N+1} \alpha \; a \right)| -  |\sin\left(\frac{\pi}{N+1} \alpha \; b \right)| \right)^2 \nonumber \\  &+& \frac{1}{2}\sum_{l \neq a} G(\epsilon_l - \epsilon_a)\frac{4\gamma^2 N}{(N+1)^2} + 
	\frac{1}{2}\sum_{l \neq b} G(\epsilon_l - \epsilon_b)\frac{4\gamma^2 N}{(N+1)^2}
\eq
We note that we always bound $G(\omega) \geq  (1+e^{4K})^{-1}$ and the two last summands can be evaluated explicitly. Disregarding the first summand, which is always 
positive, we find independent of the choice $(a,b)$  the lower bound
\bq
\Lambda_{QM} \geq \frac{4\gamma^2}{1 + e^{4K}} \frac{N(N-1)}{(N+1)^2} \geq \frac{1}{2}\frac{\gamma^2}{1+e^{4K}},
\eq
whenever $N \geq 3$. In fact, for large $N$ it is safe to state that $\Lambda_{QM} \geq 4\gamma^2(1 + e^{4K})^{-1}$. Note, that this bound is independent of the system size. We now turn to the analysis of the classical gap $\lambda_{cl}$. We observe, that the graph $E^0 = K^0$ is the complete graph since the map $S_{ab}$ induces transitions between all the different eigenstates. A direct evaluation yields 
\bq
\tau^0 \leq \max_{(a,b)} \frac{1}{P(a,b)\sigma_b} \sum_{\gamma_{ab} \ni (a,b)} \sigma_a \sigma_b |\gamma_{ab}| = \max_{(a,b)}\frac{\sigma_a}{P(a,b)},
\eq		  
Note, that since $\sigma_a = Z^{-1}e^{-K\epsilon_a} \leq N^{-1}e^{-K(\epsilon_a - \epsilon_{\max})}$, and we have that $P(a,b)$ as given above, we are left with  
\be
	\lambda_{cl} \geq 2 \gamma^2 \frac{N^2}{(N+1)^2} \geq \gamma^2,
\ee
for $N \geq 3$, since  $\tau^0 \geq \lambda^{-1}_{cl}$. For large $N$ this bound also converges to $2\gamma^2$. We conclude therefore that $\lambda \geq {\cal O}(\gamma^2)$ is independent of the system size. Together with the  mixing time bounds derived in (\ref{bound-on-t}), we therefore have that the total system equilibrates in time $t_{mix} \leq {\cal O}(\gamma^{-2}K e^{4K}\log(N\epsilon^{-2}))$.

\subsubsection{Example:  D - level system with simple transitions}
Up to now most examples could be bounded by using the simpler Gershgorin type bound stated in theorem \ref{MainBound}. We provide now an example where this bound fails and we have to make use of the more complicated bound stated in theorem \ref{robustBound}. Let us therefore consider the following system
\begin{itemize}
\item 
The Hamiltonian is assumed to be diagonal with integer eigenvalues, much like the harmonic oscillator. The $D$-level system is described by 
\be
	H = \sum_{n=1}^D \epsilon n \proj{n}.
\ee
\item
The coupling operator is now assumed to induce transitions between the different levels, however, this time we assume that the transitions are not weighted by the mode number.
In the diagonal basis of the Hamiltonian the coupling operator is of the form 
\be
	S = \gamma \sum_{n=1}^{D-1}\ket{n+1}\bra{n} + \ket{n}\bra{n+1}.
\ee
\end{itemize}

We immediately see that the set of the Bohr frequencies is identical to the Bohr frequencies of the Harmonic oscillator. They are in some sense maximally degenerate because all eigenvalues are evenly spaced. We have that $\nu = 1,\ldots,D-1$, which leads to rather large block sizes. The sets of states within each block indexed by $\nu$ are given by
\be
	\hat{\nu} = \left\{(n,n+\nu\epsilon^{-1}) | n = 1,\dots,D - \nu\epsilon^{-1} \right\}.
\ee
Both $|S_{ab}|^2 = \gamma^2(\delta_{a,b+1} + \delta_{a+1,b})$ and $|S_{ab}| = \gamma(\delta_{a,b+1} + \delta_{a+1,b})$ are readily computed. This now allows us to easily construct 
the corresponding block matrices. We immediately see that $S_{a,b} = S_{a+\nu,b+\nu}$ from which can infer that $\lambda_{-}(\underline{a},\underline{b}) = 0$. As discussed previously 
the summands that are not part of the tuples in $\hat{\nu}$ can be given by $n_1 \notin \hat{\nu}_1 = \{\nu+1,\ldots,D \}$ as well as $n_2 \notin \hat{\nu}_2 = \{1,\ldots,\nu-1\}$.\\

\begin{figure}[h]
\begin{center}
\resizebox{0.815\linewidth}{!}
{\includegraphics{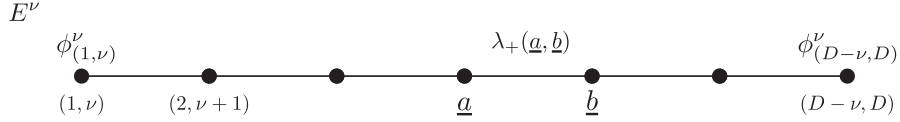}}
\caption{\label{fig:D-level} Interaction graph for the D-level system with simple transition rules. The figure depicts the transition graph $E^\nu$ for different blocks $\nu$. We observe, that due to the particular form of $S$ all $\lambda_{-}(\underline{a},\underline{b}) = 0$ vanish and we only consider the terms $\lambda_{+}(\underline{a},\underline{b}) = \gamma^2G(\epsilon_{a_1} - \epsilon_{b_1})\sigma(b_1,b_2)$.}
\end{center}
\end{figure}
Before we turn to bounding the  gap for the values $\nu > 0$,  we examine the lower bound to the classical spectral gap first. We have that $P(a,b) = (1+e^{K(b-a)})^{-1}\gamma^2$ whenever $|a - b| = 1$. 
We again make use of the bound (\ref{}), which gives 
\bq
\tau^0 &\leq& \max_{n} \frac{1}{P(n+1,n)\sigma_n} \sum_{\gamma_{ab} \ni (n,n+1)} \sigma_a \sigma_b |\gamma_{ab}|,\\
	&=&  \frac{1+e^K}{Z \gamma^2} \max_{n} e^{Kn} \sum_{a=1}^{n}\sum_{b=n+1}^D (b-a) e^{-K(a+b)},
\eq
where we have $Z = (1-e^{-K})^{-1}(e^{-K} - e^{-K(D+1)})$. We therefore can give the bound 
\be
	\tau^0 \leq \frac{1}{\gamma^2}\frac{D}{1-e^{-K}}
\ee
for the classical block. Let us know turn to the bound on $\lambda_{QM}$ as stated in theorem \ref{MainBound}. Recall, that since $\lambda_{-}(\underline{a},\underline{b}) = 0$ the sum $\sum_{\underline{n}}G(\epsilon_{n_1} - \epsilon_{m_1}) (|S_{m_1,n_1}|^2 - |S_{m_2,n_2}|^2)$ vanishes for all values of $\underline{m}$. Hence, we need to focus on the contributions arising from the  summands $\phi^\nu_{\underline{m}}$. Due to the fact that the operator $S$ only couples adjacent energy eigenstates, we have that the only contributions that arise occur at the boundaries of the graph $E^\nu$, i.e. the only terms contributing are 
\be
	\phi^\nu_{(1,\nu)} = \frac{1}{2}\frac{\gamma^2}{1 + e^K}\sigma(1,\nu) \Sp \mbox{and} \Sp \phi^\nu_{(D-\nu,D)} = \frac{1}{2}\frac{\gamma^2}{1 + e^K}\sigma(D-\nu,D).
\ee
Hence, there are many $\Lambda_{(m_1,m_2)} = 0$ so that the bound $\Lambda_{QM} = 0$ turns out to be useless. However, we can try to evaluate the bound $\hat{\Lambda}_{QM}$. Observe, that since we only have two summands $\phi^\nu_{\underline{m}}$ which do not vanish and furthermore we have that $\lambda_{-}(\underline{a},\underline{b}) = 0$ the variables 
\bq
\omega_{\underline{m}}^\nu(\underline{a},\underline{b}) = \left \{\begin{array}{l} \phi^\nu_{(1,\nu)} \;\;\Sp:\Sp \mbox{for} \Sp b_2 \geq m_1 \nonumber 
\\ \phi^\nu_{(D-\nu,D)} : \Sp \mbox{for}  \Sp b_1 \leq m_2  \end{array}\right.,
\eq
are readily constructed, since $E^\nu$ is a tree already. Moreover, the normalization $N_\nu = \phi^\nu_{(1,\nu)} + \phi^\nu_{(D-\nu,D)}$ follows immediately. Hence, we can bound 
the Frobenius norm $W^\nu$ immediately and obtain
\bq
\hat{\Lambda}_\nu^{-1} &=& \sum_{\underline{m} \in \hat{\nu}}\frac{\sigma(m_1,m_2)}{4N_\nu^2}\left(\phi^\nu_{(1,\nu)} + \phi^\nu_{(D-\nu,D)}  \sum_{(\underline{l},\underline{n}) \in E^\nu}  \frac{\left[\omega^\nu_{\underline{m}}(\underline{l},\underline{n})\right]^2 }{\lambda_{+}(\underline{l},\underline{n})} \right) \\
&\leq& \frac{1}{\gamma^2} \frac{e^K}{1-e^{-K}}\left(1 + D - \nu \right)   
\eq
Hence, choosing $\nu = 1$ gives rise to  
\be
\hat{\Lambda}_{QM} \geq \gamma^2e^{-K}\left(1 - e^{-K}\right)\frac{1}{D},
\ee 
and thus the total lower bound to gap $\lambda$ of the map. Hence, when applying the bound on the mixing time, we obtain $t_{mix} = {\cal O}(\gamma^{-2}e^KD^2\log(\epsilon^{-2}))$. Note, that the behavior of the spectral gap changes in the infinite temperature limit and one obtains a scaling $\lambda \approx {\cal O}(D^{-2})$.

\section{Conclusions}

We have constructed a method to find lower bounds to the spectral gap of a family of Lindblad  operators which are known as Davies generators. The formalism developed can be seen as a generalization of the the canonical paths method for classical Markov processes. The spectral gap for reversible semi-groups can immediately be used to estimate the convergence, or mixing time of the semi group by methods derived in \cite{chi2}. An important property of Davies generators is that the matrix associated to the Dirichlet form decomposes into orthogonal blocks when working in the eigenbasis of the Hamiltonain. One can therefore bound the spectral gap in each of the smaller sub problems individually and then choose the smallest value as the full lower bound of the gap. We have seen that this block structure becomes particularly simple, when the eigenvalues of the system Hamiltonian are non-degenerate. In particular, one finds that a single block corresponds to the dynamics of the Pauli master equation \cite{Davies,Davies2,alicki1977markov} for the diagonal entries of the density matrix. For this block well known classical techniques can be applied directly. However, looking only at the classical spectral gap does not suffice to give estimates on the total spectral gap, as a simple counter example has shown. We had to derive new bounds on the smallest eigenvalues in each of the other blocks in order to provide an actual lower bound to the spectral gap of the generator. \\

One general restriction of this approach is the fact that we need to require that the spectrum of the system Hamiltonian is non-degenerate. One could assume that this is the case for many naturally occurring systems, in particular when perturbations are present that lift the degeneracies. However, it would nevertheless be very interesting, in particular in light of applications to the estimation of survival times of passive quantum memories, if one could derive similar bounds in the presence of degeneracies. It is conceivable, that the approach taken here can be directly applied to some example systems when one doesn't attempt to derive a generic bound as we have done here. One can observe, that the block structure of lemma \ref{non-degForm} prevails even in the presence of degenerate eigenvalues. This can be seen by simply assuming that the projectors $\Pi_m$ are supported on the degenerate subspaces. We see that eqn. (\ref{blockDeg}) then gives rise to a similar block structure as stated in the lemma. This could be a natural starting point for an attempt to derive bounds also for degenerate system Hamiltonians. 

\vspace{1cm}
\textbf{Acknowledgements}
I would like to thank Michael J. Kastoryano and Fernando Pastawski for fruitful discussions. In particular I would like to thank Fernando Pastawski for pointing out a mistake in a previous counter example. The author gratefully acknowledges the support from the Erwin Schr\"odinger fellowship, Austrian Science Fund (FWF): J 3219-N16.

\bibliographystyle{unsrt}

\end{document}